\documentclass[twocolumn
 ,secnumarabic,amssymb, amsmath,nobibnotes, aps, prl,superscriptaddress]{revtex4}
\usepackage{docs}%
\usepackage{bm}%
\usepackage{graphicx}% Include figure files
\begin{document}

\title{Intermittent permeation of cylindrical nanopores by water}

\author{Rosalind Allen}%
\email{rja22@cam.ac.uk}
\affiliation{Department of Chemistry, University of Cambridge, Lensfield Road, Cambridge CB2 1EW, United Kingdom}

\author{Simone Melchionna}%

\affiliation{Istituto per le applicazioni del calculo  ``M. Picone'', Consiglio Nazionale delle Ricerche, V. le del Policlinico 137, 00161 Rome, Italy }

\author{Jean-Pierre Hansen}%

\affiliation{Department of Chemistry, University of Cambridge, Lensfield Road, Cambridge CB2 1EW, United Kingdom}

\date{\today}%

\begin{abstract}
Molecular Dynamics simulations of water molecules in nanometre sized cylindrical channels connecting two reservoirs show that the permeation of water is very sensitive to the channel radius and to electric polarization of the embedding material. At threshold, the permeation is {\emph{intermittent}} on a nanosecond timescale, and strongly enhanced by the presence of an ion inside the channel, providing a possible mechanism for gating. Confined water remains surprisingly fluid and bulk-like. Its behaviour differs strikingly from that of a reference Lennard-Jones fluid, which tends to contract into a highly layered structure inside the channel.  
\end{abstract}

\maketitle

Highly confined molecular or ionic fluids, including fluid films compressed 
by surface force machines \cite{REF1}, or trapped in nanoporous materials 
like zeolites \cite{REF2}  or carbon nanotubes \cite{REF3,REF4}, exhibit 
distinctive structural, dynamical and phase behaviour which may differ 
widely from their corresponding bulk properties. Recent experiments \cite{klein} 
and simulations \cite{REF4} %strongly suggest the very specific
demonstrate the interesting behaviour 
of water under conditions of extreme confinement, which 
%may be traced back 
is related to the disruption of the hydrogen-bond network between H$_2$O molecules, 
depending on the hydrophobic or hydrophilic nature of the confining surface. 
%An understanding of the molecular organization of
A good understanding of the molecular organization of confined water is essential %to gain insight into 
to the study of
the selectivity and permeation mechanisms of 
ion channels through membranes \cite{REF6}.\par
While much of the published simulation work is on realistic or semi-realistic 
models of specific ion channels \cite{REF7}, the present letter examines 
the generic behaviour of water permeation in a highly simplified model of a 
channel, as a function of a small number of physical parameters characterizing 
the channel% radius and length
. The proper inclusion of end effects and 
Coulombic interactions increases the relevance of the present results for 
real ion channels, compared to earlier studies of infinite 
channels \cite{REF8,REF9,allen}. \par
\begin{figure}
\scalebox{0.5}{\includegraphics[trim=0 50 10 100,clip=true]{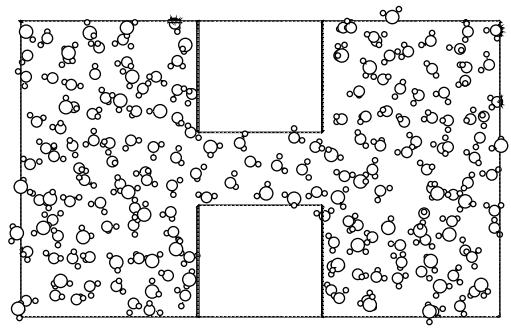}}
\caption{\label{fig1}Schematic representation of a cut through the simulation cell. The $z-$axis coincides with the axis of the cylindrical channel connecting the two reservoirs; the interior wall region is a continuum of permittivity $\epsilon$. Periodic boundary conditions apply in all three directions. The channel radius and length are shown at their effective values, $R_{eff} < R$ and $L_{eff} > L$, as explained in the text.}
\end{figure}

The model considered in the present work is that of a cylindrical channel of 
radius $R$ and length $L$ through a slab of dielectric material, of relative 
permittivity $\epsilon$ (representing the membrane), which separates two reservoirs containing water 
molecules and eventually ions (representing the intra and extra-cellular 
compartments), symmetrically placed on both sides of the 
slab% representing the membrane separating intra and extra-cellular compartments 
 (see Fig.~\ref{fig1}). In our Molecular Dynamics (MD) simulations\cite{dlprotein}, the 
two reservoirs, which are connected through a common x-y plane under periodic 
boundary conditions, contain typically $N=500$ water molecules each. The 
%average water density 
number of water molecules in the reservoirs fluctuates somewhat when %water molecules enter or exit the cylindrical channel
the cylindrical channel fills or empties; in order to maintain a 
constant density of $0.996$~gr~cm$^{-3}$, the reservoir length parallel to the axis of the channel (taken to be $z$) and the z-components of the molecular position vectors are 
scaled regularly using a version of the Berendsen barostat \cite{berendsen}.\par

The water molecules interact via the standard SPC/E pair potential 
\cite{REF10} which involves a Lennard-Jones (LJ) potential between the 
O-atoms ($\sigma= 0.3169$~nm, $\epsilon=0.6502$~kJ~mol$^{-1}$ i.e $\epsilon / k_B= 78.2$~K ), and Coulombic
interactions between the sites associated with the O atoms ($q=-0.8476$) and 
the two H atoms ($q=+0.4238$). The simulations reported in this letter 
contain no  ions or just one K$^+$Cl$^-$ pair, which interact with the 
water molecules as in ref. \cite{REF8}. The radial interaction between the 
channel surface and the water molecules is taken to be the LJ potential 
between a CH$_3$ group and an O atom \cite{REF11}, integrated over the 
surface of an infinitely long cylinder \cite{REF12}.  The same potential, 
now as a function of the distance from the planar wall, is used for 
the membrane-water interaction, with an appropriate rounding and shifting procedure to 
avoid any discontinuities. A similar procedure is used for ion-surface interactions.\par
Since water molecules and ions carry charged sites, they polarize the 
membrane modelled here by the dielectric slab. The interface between slab, 
channel or reservoir is treated as infinitely sharp (dielectric discontinuity),
so that the polarization charge is entirely localized on 
the surface of the slab and can be calculated by the variational procedure 
described in ref. \cite{REF13}; the numerical implementation requires the 
use of a grid spanning the cylindrical surface and the planar interface 
between the slab and the reservoirs. Most of the results presented in this 
letter neglect membrane polarization, i.e. $\epsilon=1$ is assumed; the 
importance of polarization on the behaviour of confined water is assessed 
later in the letter, by comparing data obtained for $\epsilon = 1$ and 
$\epsilon=4$. \par
The present work is restricted to equilibrium fluctuations. Starting from 
initial water configurations with an empty channel, the number of molecules 
within the channel is monitored as a function of time, for total times up 
to $6.5$~ns. Simulations were carried out for a rather short channel of 
length $L=0.8$~nm. The channel radius $R$ was varied between $0.4$ 
and $0.7$~nm, but in view of the soft nature of the surface-water 
interaction, the corresponding effective radii of the cylinders accessible 
to the centres of the O atoms (i.e. such that the wall-O repulsive potential is less than $k_BT$) are $R_{eff}= 0.17 - 0.47$~nm. Similarly the effective length of the channel is $L_{eff} = 1.26$~nm, which is roughly the length of the selectivity filter in the KcsA channel, the structure of which has recently been resolved by X-ray diffraction \cite{doyle,REF14}.\par
\begin{figure}
\scalebox{0.25}{\rotatebox{0}{\includegraphics{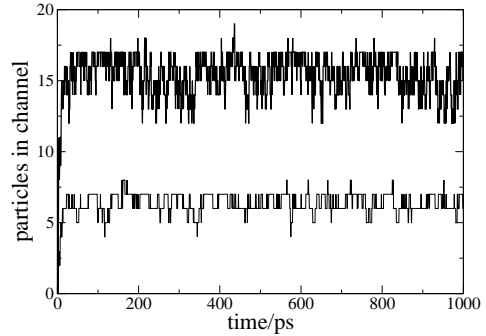}}}
\caption{\label{fig2} Number $n(t)$ of OW atoms inside channels of radius $R=0.45$~nm ($R_{eff}=0.22$~nm) and $R=0.6$~nm ($R_{eff}=0.37$~nm), in bold, as a function of time (in ps).}
\end{figure}

When the channels are filled, the radial density profiles of O and H atoms 
can be calculated by averaging over all water configurations generated in the 
MD runs. In order to gain a better understanding of the importance of the 
hydrogen-bond network within the channel, simulations were also carried out 
for a reference system of non-associative molecules obtained by removing 
the H atoms, i.e. an imaginary system of ``OW'' atoms interacting solely through 
the Lennard-Jones part of the SPC/E potential. While the H$_2$O simulations were run at room temperature, 
the reference runs for the OW atoms were carried out at $T=120$~K and a reduced 
density $\rho^\star=n\sigma^3/V=0.85$, conditions under which the bulk LJ system 
is expected to be liquid. Under these conditions we also find liquid-like diffusion in the pore, whereas at the lower temperatures $T=56.2$~K (i.e. $T^*=k_BT/\epsilon = 0.72$, close to the LJ triple point\cite{REF15}) and $T=85$~K the OW atoms form a disordered solid in the pore.  The number $n(t)$ of OW atoms 
within channels of two different radii are shown in Fig.~\ref{fig2} as a function 
of time. As expected, there are large fluctuations in $n(t)$, but the key 
observation is that the LJ particles fill the channel for all radii 
considered. The mean density of particles inside the channel, 
$\langle n(t)\rangle/(\pi R_{eff}^2 L)$ is about 1.7 times larger than the 
bulk density in the reservoirs, but the confined atoms remain in a fluid 
state, as characterized by a diffusion constant along the cylinder axis ($2\times 10^{-5}$~cm$^2$~s$^{-1}$) comparable to the bulk diffusion constant ($6\times 10^{-5}$~cm$^2$~s$^{-1}$). The radial density 
profile $\rho(r)$ is highly structured, as shown in Fig. 4d, indicative 
of layering.\par
\begin{figure}
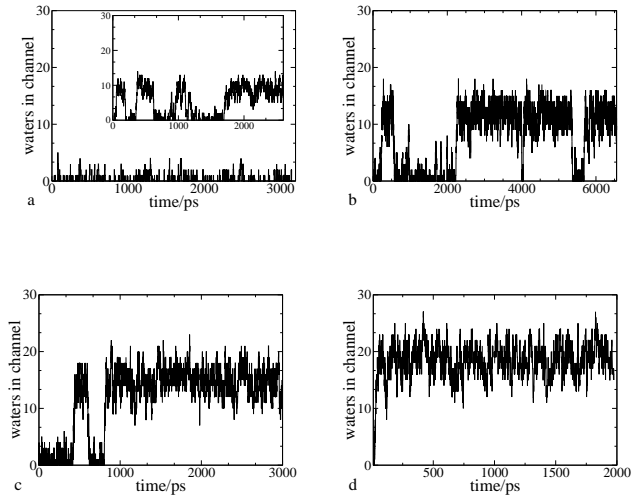

\scalebox{0.15}{\rotatebox{0}{\includegraphics{fig3a.eps}}}\hspace{0.5cm}
\scalebox{0.15}{\rotatebox{0}{\includegraphics{fig3b.eps}}}\vspace{1cm}
\scalebox{0.15}{\rotatebox{0}{\includegraphics{fig3c.eps}}}\hspace{0.5cm}
\scalebox{0.15}{\rotatebox{0}{\includegraphics{fig3d.eps}}}
\caption{\label{fig3} Number $n(t)$ of water molecules inside channels of length $L=0.8$~nm ($L_{eff}=1.26$~nm) and radii $R=0.55$~nm ($R_{eff}=0.32$~nm) (a), $R=0.6$~nm ($R_{eff}=0.37$~nm) (b), $R=0.65$~nm ($R_{eff}=0.42$~nm) (c) and $R=0.7$~nm ($R_{eff}=0.47$~nm) (d) as a funtion of time (in ps); note that the total durations of the four runs vary; all four runs are for $\epsilon=1$; the inset in frame (a) shows $n(t)$ for the same $L$ and $R$, but with $\epsilon=4$.}
\end{figure}

The behaviour changes dramatically when the OW atoms are replaced by 
H$_2$O molecules. The numbers of molecules inside the channel are shown 
as a function of time in Fig. ~\ref{fig3} for four different cylinder radii. 
When $R=0.55$~nm, the channel shows no sign of filling and the small fluctuations 
in $n(t)$ away from zero reflect attempts by small numbers of molecules 
to penetrate the channel at either end. At the slightly larger channel 
radius $R=0.6$~nm, a strikingly intermittent behaviour is apparent, 
reminiscent of recent observations in simulations  of water in carbon 
nanotubes \cite{REF4} and of water in atomistically rough models of ion 
channels \cite{REF6}. States where the channel is filled alternate 
stochastically with empty channel states, on a time scale of typically 1ns. 
When the radius is further increased to $R=0.65$~nm, the channel appears 
to remain filled after an initial intermittency, although it cannot be ruled 
out that short periods in the empty state could be observed in much 
longer runs. Finally at $R=0.7$~nm, the channel appears to remain filled 
throughout, with a mean density of $0.034$~\AA$^{-3}$ comparable to the bulk density 
in the reservoirs, $0.033$~\AA$^{-3}$. Diffusion of the confined molecules along the $z-$axis 
occurs on a timescale comparable to bulk diffusion (the reservoir diffusion constant and the channel $z-$axis diffusion constant both being  $3\times 10^{-5}$~cm$^2$~s$^{-1}$).
Radial density profiles of the O and H atoms, averaged over occupied 
states of the channel, are shown in Fig.4 a-c for three values of the 
radius $R$. They are surprisingly flat, and contrast sharply with the density 
profile of the Lennard-Jones particles plotted in Fig. 4d. The absence 
of a clear-cut layering of the H$_2$O molecules, as well as the substantial 
fluidity of the confined water, characterized by the bulk-like diffusion, 
agree qualitatively with recent surface force apparatus measurements of 
highly confined water films\cite{klein}.\par

The results in Fig.~\ref{fig3} show that the filling (or wetting) of a 
cylindrical channel is very sensitive to its radius, with a threshold 
radius below which water does not penetrate on this timescale.  So far 
the polarization of the confining protein and membrane has been 
neglected, by setting $\epsilon=1$. One intuitively expects that 
such polarization by the charge distribution on the water 
molecules will lower the electrostatic energy, and hence favour the 
wetting of narrow channels. This is confirmed by the time trace of $n(t)$ 
shown in the inset to Fig.~\ref{fig3}a from MD simulations with $\epsilon = 4$: water molecules 
are seen to fill the $R=0.55$~nm channel intermittently, while the same 
channel remains empty in the absence of membrane polarization. The latter 
effect is thus found to lower the threshold radius above which water 
molecules can penetrate inside the channel. The effect is expected to be 
significantly enhanced for larger values of $\epsilon$, which may be more 
appropriate for the channel protein and membrane.\par

\begin{figure}
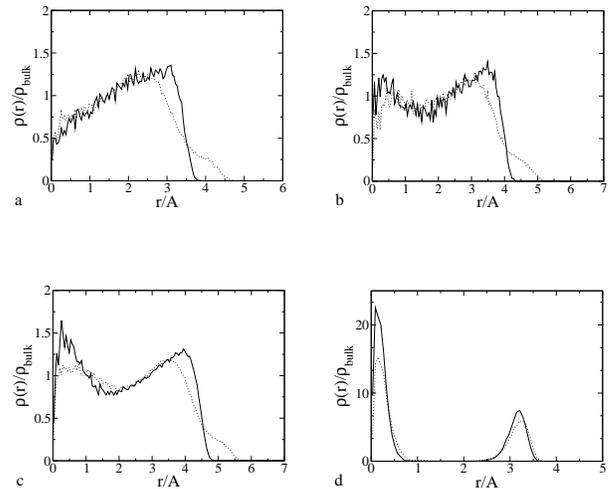

\scalebox{0.15}{\rotatebox{0}{\includegraphics{fig4a.eps}}}\hspace{0.5cm}
\scalebox{0.15}{\rotatebox{0}{\includegraphics{fig4b.eps}}}\vspace{1cm}
\scalebox{0.15}{\rotatebox{0}{\includegraphics{fig4c.eps}}}\hspace{0.5cm}
\scalebox{0.15}{\rotatebox{0}{\includegraphics{fig4d.eps}}}
\caption{\label{fig4} Radial density profiles $\rho_{\alpha}(r)$ of O atoms (full curves) and H atoms (dotted curves) of water in channels of length $L=0.8$~nm ($L_{eff}=1.26$~nm) and radii $R=0.6$~nm ($R_{eff}=0.37$~nm) (a), $R=0.65$~nm ($R_{eff}=0.42$~nm) (b) and $R=0.7$~nm ($R_{eff}=0.47$~nm) (c). Averages are taken only over times when the channel is filled. Frame (d) shows the radial profiles for OW atoms in a channel of the same length and  $R=0.6$~nm ($R_{eff}=0.37$~nm), at temperatures $T=85$~K (dotted curve) and $T=120$~K (full curve).}
\end{figure}

Returning to the non-polarizable case $\epsilon=1$, water can be made to 
fill a narrow channel by placing an ion inside. In an MD simulation a $K^+$ 
ion was held fixed on the axis in the middle of a channel of radius 
$R=0.5nm$; the channel was then found to be rapidly filled by H$_2$O 
molecules with an average occupation number $n \approx 6.5$. Water molecules were found to penetrate channels even with $R$ as small as $0.3$~nm under these circumstances.\par 

A close examination of molecular configurations, radial number density 
profiles and mean occupation numbers leads to the conclusion that water 
fills ion channels beyond a critical, $\epsilon$-dependent radius in a 
surprisingly uniform manner which is intermittent near the threshold. 
As soon as water wets the channel, the water molecules appear 
to fill the available volume , with a mean density 
comparable to the bulk density of liquid water. In this series of simulations, no long-lived single-file chains of water molecules were observed. Using an energetic 
criterion\cite{jorgensen} the mean number of hydrogen bonds was found 
to be $2.5$ in a channel of $R=0.7$~nm, compared to $3.5$ in the reservoir.\par 

In summary we have investigated a highly simplified, generic model 
for the selectivity filter of ion channels, which is entirely 
characterized by three physical parameters $R$, $L$ and $\epsilon$, once one has fixed the interaction parameters of the wall. The present investigation 
was limited to $L=0.8$~nm. At the threshold radius, water fills the channel 
intermittently, and the fraction 
of time over which the channel is in the occupied state depends sensitively 
on $R$ and $\epsilon$. 
%The intermittency ``periods'', i.e. the mean durations of the filled and empty states, are expected to increase channel length $L$, because the relative weight of fluctuations at tboth ends will then decrease. (I'M NOT SURE THE LENGTH OF BOTH FILLED AND EMPTY STATES WILL INCREASE) 
The timescale for the intermittency is expected to increase with channel length $L$, because larger fluctuations at the channel ends will be required for channel filling.
Finite, physiologically relevant concentrations 
of salt will be included in future work. In view of the preliminary 
results with one ion fixed within the channel, it is expected that 
permeation of narrow channels by water molecules and ions is a 
very cooperative effect, where the polarization of the channel surface 
will play a crucial role, which may provide a simple scenario for ion 
channel gating.\par

\begin{acknowledgments}
The authors acknowledge constructive discussions with Ruth Lynden-Bell and Bob Eisenberg. RA is grateful to the EPSRC and to Unilever for a post-graduate CASE award and SM is grateful to the Leverhulme Trust for post-doctoral support during the early stages of this project.
\end{acknowledgments}

\end{document}